\documentclass[prc,onecolumn]{revtex4}
\usepackage{amsmath,latexsym,amssymb}
\usepackage[dvips]{graphicx}
\usepackage{epsfig}

\def\vec#1{\mathchoice
 {\mbox{\boldmath $\displaystyle#1$}}
 {\mbox{\boldmath $\textstyle#1$}}
 {\mbox{\boldmath $\scriptstyle#1$}}
 {\mbox{\boldmath $\scriptstyle#1$}}}

\begin{document}

\title{Relativistic Hartree approach with
exact treatment of vacuum polarization for finite nuclei}

\author{
Akihiro Haga$^1$\footnote{Electronic address:
haga@rcnp.osaka-u.ac.jp}, 
Setsuo Tamenaga$^1$\footnote{Electronic address:
stame@rcnp.osaka-u.ac.jp}, 
Hiroshi Toki$^1$\footnote{Electronic
address: toki@rcnp.osaka-u.ac.jp}, and 
Yataro Horikawa$^2$\footnote{Electronic address:
horikawa@sakura.juntendo.ac.jp} }

\affiliation{
$^1$Research Center for Nuclear Physics (RCNP), Osaka University,
Ibaraki, Osaka 567-0047, Japan
\\
$^2$Department of Physics, Juntendo University, Inba-gun, Chiba
 270-1695, Japan }
%$^3$ Department of  Environmental Technology and Urban Planning,
%Nagoya Institute of Technology, Gokiso, Nagoya 466-8555, Japan }

\date{25 August. 2004}

\setlength{\baselineskip}{0.3in}

%\maketitle

\begin{abstract}
\setlength{\baselineskip}{0.3in}

We study the relativistic Hartree approach with the exact
treatment of the vacuum polarization in the Walecka
($\sigma-\omega$) model. The contribution from the vacuum
polarization of nucleon-antinucleon field to the source term of
the meson fields is evaluated by performing the energy integrals
of the Dirac Green function along the imaginary axis. With the
present method of the vacuum polarization in finite system, the
total binding energies and charge radii of $^{16}$O and $^{40}$Ca
can be reproduced. On the other hand, the level-splittings in the
single-particle level, in particular the spin-orbit splittings,
are not described nicely because the inclusion of vacuum effect
provides a large effective mass with small meson fields. We also
show that the derivative expansion of the effective action 
which has been used to calculate the vacuum contribution 
for finite nuclei gives a fairly good approximation.

PACS number(s): 21.10.-k,21.60.-n, 13.75.Cs
\end{abstract}

\maketitle

\section{Introduction}

It is well known that the relativistic field theory based on the
hadrons referred to as quantum hadrodynamics (QHD) has been of
great success in describing the ground states of finite
nuclei\cite{WA74}. When the energy functional of the relativistic
mean field (RMF) is fitted to nuclear saturation, the RMF model
automatically produces an appropriate order of the spin-orbit
splitting of nuclei, the spin-observables of the proton-nucleus
scattering and energy dependence of the proton-nucleus optical
potential. In RMF, however, only positive-energy nucleons are
taken into account in the calculation in spite of the existence of
solutions with negative energy, which is one of the interesting
characters of the relativistic picture. The negative-energy states
are observables in antinucleonic atom.
The bound levels of antinucleonic atom are
predicted to be much deeper than those of ordinary nucleon since
the magnitudes of nucleon-scalar and -vector self energies, which are
very large, cancel each other to provide the usual binding
energy in the nucleon sector, while they add each other in the
antinucleon sector (see Ref.~\cite{KL02} and references therein).
It is important to study if this feature remains to be the case, 
when we treat the negative-energy states explicitly in the QHD framework. 
In particular, to know how deep the antinucleons are
bound in the nucleus provides valuable
information for the search of the exotic 
collective antimatter production proposed by Greiner\cite{GR95}.

Recently, the gauge invariant nuclear-polarization calculation was
carried out in the relativistic random-phase approximation (RRPA)
based on the RMF theory.
It was found that the RRPA eigenstates with negative energy have 
a significant role because the transverse form factors to these 
states become considerably large\cite{HA04}. 
If antinucleons are deeply bound, the transverse response
function has a contribution from the antinucleon states
with lower energy than twice the nucleon mass. Then,
electron-scattering or photo-absorption may confirm the large
binding of antinucleon. In order to investigate the antinucleon
state in the QHD model, however, it is required to take into
account the effect of vacuum polarization of nucleon-antinucleon
field to the mean field because of consistency. Namely, the RMF
model without the vacuum contribution has to be extended to the
full one-nucleon-loop approximation, which we refer to as the
relativistic Hartree approximation (RHA).

The vacuum contributions and their effects on the bound states of
positive energy were investigated by several authors within the
local-density approximation and the derivative-expansion method
\cite{HO84,PE86,IC88,WA88,FO89,FU89,MA99,MA03}. After refitting
the parameters of the model to the properties of spherical nuclei,
it was found that the RHA calculation can reproduce the
experimental data of the binding energies and the rms radii of nuclei 
as well as the RMF can. On the other hand, 
due to the decrease of the scalar and vector
potentials by the feedback from the vacuum in the RHA calculation, 
the effective mass of nucleon  
becomes large and the binding energy of
antinucleon becomes small compared with the RMF.

Despite the finding of the importance of the vacuum effect,
however, the exact evaluation of one-loop corrections in a finite
nuclear system has never been performed. This is an exceedingly
difficult task, since the exact treatment of vacuum polarization
in finite nuclei requires the computation of not only the valence
nucleon with positive energy but also infinite number of the
negative-energy states. In this context, there is an excellent
method developed in the quantum electrodynamics (QED)
\cite{GY72,SO88}; the summation of the eigenstates is replaced by
the energy integral of the Dirac Green function along the
imaginary axis in which the Green functions do not oscillate and
therefore, it is possible to calculate it much faster. 
Blunden carried out the exact RHA calculation of QHD with 1+1 dimensions and
the calculation of quantum solitons model with 3+1 dimensions\cite{BL90,ST97}.
In the present paper we will apply this method developed in atomic
physics to the RHA calculation of QHD.

This paper is organized as follows. In Sec.~II.A
we introduce the effective Lagrangian density used in this work
and give the outline of the renormalization in the source term of
meson fields.
The method to obtain the renormalized vacuum densities
is given in Sec.~II.B for baryon density,
and in Sec.~II.C for scalar density.
The detail of the computational procedure in the
calculation of vacuum correction
are described in Sec.~III.
In Sec.~IV, the numerical results of the RHA calculation including
vacuum corrections for baryon and scalar densities
are presented for $^{16}$O and $^{40}$Ca and the role of vacuum
in the properties of nucleus is discussed.
In Sec.~V, we will also compare our rigorous method
with the local-density approximation and the derivative expansion.
Finally we give summary of the present calculation in Sec.~VI.

\section{Relativistic Hartree Approach in Finite Nucleus}
\subsection{Lagrangian density}

The nucleus is described as a
system of Dirac nucleons which interact in a relativistic
covariant manner through the exchange of several mesons;
scalar meson ($\sigma$) produces a strong attraction while
isoscalar vector meson ($\omega$) produces a strong repulsion
for the nucleon sector.
In the present work,
we employ the Lagrangian density including photon ($A$)
as well as $\sigma$ and $\omega$ mesons as
\begin{eqnarray}
{\cal L}_N &=&
\bar{\psi}_N
(i\gamma^{\mu}\partial_{\mu}-m_N)
{\psi}_N
+\frac{1}{2}\partial_{\mu}\sigma\partial^{\mu}\sigma
-\frac{1}{2}m_{\sigma}^2\sigma^2
-\frac{1}{3}g_{2}\sigma^3
-\frac{1}{4}g_{3}\sigma^4
\nonumber\\
&&
-\frac{1}{4}(\partial_{\mu}\omega_{\nu}-\partial_{\nu}\omega_{\mu})^2
+\frac{1}{2}m_{\omega}^2\omega_\mu\omega^\mu
\nonumber \\
&&
-\frac{1}{4}(\partial_{\mu} A_{\nu}-\partial_{\nu} A_{\mu})^2
\nonumber \\
&&
-g_\sigma \bar{\psi}_N \sigma {\psi}_N
-g_\omega \bar{\psi}_N \gamma_\mu\omega^\mu {\psi}_N
-\frac{1}{2}e \bar{\psi}_N (1+\tau_3)\gamma_\mu A^\mu {\psi}_N
-\delta {\cal L},
\label{nucleusL}
\end{eqnarray}
where
\begin{eqnarray}
\delta {\cal L}=-\frac{1}{4}\zeta_\omega
(\partial_{\mu}\omega_{\nu}-\partial_{\nu}\omega_{\mu})^2
+\frac{1}{2}\zeta_\sigma\partial_{\mu}\sigma\partial^{\mu}\sigma
+\alpha_1\sigma
+\frac{1}{2}\alpha_2\sigma^2
+\frac{1}{3}\alpha_3\sigma^3
+\frac{1}{4}\alpha_4\sigma^4
\label{nucleusDL}
\end{eqnarray}
denotes counterterms to renormalize the nucleon density which has
the divergence from the vacuum. Since this Lagrangian density
includes the nonlinear coupling terms for $\sigma$ meson, the
one-$\sigma$-meson loop also can contribute to the nucleon
density\cite{FO89,MA99}. In the present paper, however, we neglect
this contribution for simplification. 
Assuming that
the nuclear ground state is spherically symmetric, the Hartree
basis consists of eigenfunctions of the following Dirac equation;
\begin{equation}
\left[\gamma^0(\epsilon_N-g_\omega\omega_0(\vec{r})
-\frac{1}{2}e(1+\tau_3)A_0(\vec{r}))+i\vec{\gamma}\cdot
\vec{\nabla}-(m_N+g_\sigma\sigma(\vec{r}))
\right]\psi_N(\vec{r})=0.
\label{Dirac}
\end{equation}
The meson fields $\omega_0(\vec{r})$ and $\sigma(\vec{r})$
satisfy the Klein-Gordon equations,

\begin{eqnarray}
(-\nabla^2+m_{\omega}^2)\omega_0(\vec{r})&=& g_\omega 
\rho_{\omega\hspace{1mm}{\rm ren}}(\vec{r}),
\\
(-\nabla^2+m_{\sigma}^2)\sigma(\vec{r})&=& - g_s
\rho_{\sigma\hspace{1mm}{\rm ren}}(\vec{r})
-g_2\sigma^2(\vec{r})
-g_3\sigma^3(\vec{r}),
\end{eqnarray}
respectively.
The renormalized baryon ($\rho_{\omega\hspace{1mm}{\rm ren}}(\vec{r})$)
and scalar ($\rho_{\sigma\hspace{1mm}{\rm ren}}(\vec{r})$) densities
are given by
\begin{eqnarray}
\rho_{\omega\hspace{1mm}{\rm ren}}(\vec{r})&=& 
\int_{C}\frac{dz}{2\pi i}{\rm Tr}[\gamma_0 G^H(\vec{r},\vec{r};z)]
%+c_3\omega^3_0(\vec{r})
+({\rm CT})
\label{KGw}
\\
\rho_{\sigma\hspace{1mm}{\rm ren}}(\vec{r})
&=&
\int_{C}\frac{dz}{2\pi i}{\rm
Tr}[G^H(\vec{r},\vec{r};z)]+({\rm CT}), 
\label{KGs}
\end{eqnarray}
where $G^H(\vec{r},\vec{r};z)$ is single-particle
Green function of the Hartree approximation 
with the potential terms for the Dirac equation
(\ref{Dirac}). The $z$ integrations are carried out along the
modified Feynman contour which is below the real axis up to
nuclear Fermi energy\cite{GY72,SO88}.  
The divergences arising from these integrals
are removed by the contributions from the counterterms denoted by
CT. This procedure will be discussed in detail in subsection II.B
and II.C.
The integral along the Feynman contour 
can be changed to the integral over the imaginary $z$ axis
with the additional pole contribution of 
the positive-energy states up to Fermi level.
Thus, we may write unrenormalized baryon and scalar densities as
\begin{eqnarray}
\int_{C}\frac{dz}{2\pi i}{\rm Tr}[\gamma^0G^H(\vec{r},\vec{r};z)]
&=& \sum_{\epsilon_i>0}^{F} \psi_i^\dagger(\vec{r})\psi_i(\vec{r})
- \int_{-i\infty}^{+i\infty}\frac{dz}{2\pi i}{\rm Tr}[\gamma^0G^H_V(\vec{r},\vec{r};z)]
\label{rhow}\\
\int_{C}\frac{dz}{2\pi i}{\rm Tr}[G^H(\vec{r},\vec{r};z)]
&=& \sum_{\epsilon_i>0}^{F} \bar{\psi}_i(\vec{r})\psi_i(\vec{r})
+ 
\int_{-i\infty}^{+i\infty}\frac{dz}{2\pi i}{\rm Tr}[G^H_V(\vec{r},\vec{r};z)]
\label{rhos},
\end{eqnarray}
respectively. Here, $G^H_V$ is the vacuum part of the single-particle Green function 
of the relativistic Hartree approximation:
\begin{eqnarray}
G^H(\vec{r}_1,\vec{r}_2;z) &=& G^H_D(\vec{r}_1,\vec{r}_2;z) + G^H_V(\vec{r}_1,\vec{r}_2;z)
\\
G^H_D(\vec{r}_1,\vec{r}_2;z) &=& 2\pi i \sum_{\epsilon_i>0}^{F}\delta(z-\epsilon_i)
\psi_i(\vec{r_1})\bar{\psi}_i(\vec{r_2})
\\
G^H_V(\vec{r}_1,\vec{r}_2;z) &=& \sum_{\epsilon_i>0}
\frac{\psi_i(\vec{r_1})\bar{\psi}_i(\vec{r_2})}{z-\epsilon_i+i\eta}
+\sum_{\epsilon_i<0}
\frac{\psi_i(\vec{r_1})\bar{\psi}_i(\vec{r_2})}{z-\epsilon_i-i\eta}
\end{eqnarray}
The numerical integration for $G^H_V$ along the
imaginary $z$ axis can be carried out straightforwardly because
there are no poles in the integrand. Although the second terms of
right-hand side of (\ref{rhow}) and (\ref{rhos}) have divergence, an expansion
of the total vacuum correction in the coupling constants
$g_\omega$ and $g_\sigma$ of meson fields verifies that all
divergences are contained in the first order of $g_\omega$ for
baryon density, and are contained up to the third order of
$g_\sigma$ for scalar density. In the following subsections, we
show that these divergences are removed by taking the proper
counterterms (\ref{nucleusDL}) into account.

\subsection{Vacuum Correction for Baryon Density}

In this subsection, we consider the vacuum correction
for baryon density. For the estimation of vacuum correction,
we will treat proton and neutron on the equal footing to save numerical effort.
In order to deal with the unrenormalized density containing the divergence,
we start with the perturbative expansion by $\omega$ and $\sigma$ fields;
\begin{eqnarray}
\int_{-i\infty}^{+i\infty}\frac{dz}{2\pi i}{\rm Tr}[\gamma^0G^H_V(\vec{r},\vec{r};z)]
&=&
\int_{-i\infty}^{+i\infty}\frac{dz}{2\pi i}{\rm Tr}[\gamma^0G^0(\vec{r},\vec{r};z)]
\nonumber \\
&&
+ \int_{-i\infty}^{+i\infty}\frac{dz}{2\pi i}{\rm Tr}[\int d\vec{x}
\gamma^0 G^0(\vec{r},\vec{x};z) g_\omega 
\gamma^0 \omega(\vec{x})G^0(\vec{x},\vec{r};z)]
\label{xpanw}\\
&&
+ \int_{-i\infty}^{+i\infty}\frac{dz}{2\pi i}{\rm Tr}[\int d\vec{x}
\gamma^0 G^0(\vec{r},\vec{x};z) g_\sigma 
\sigma(\vec{x})G^0(\vec{x},\vec{r};z)]
\nonumber \\
&&+ {\rm higher} \hspace{1mm} {\rm order} \nonumber
\end{eqnarray}
where $G^0$ denotes the Green function for the free Dirac equation. 
This expansion allows us to write the vacuum polarization 
as a sum of infinite terms as shown in Fig.~1.

%\vspace{30mm}

We note that only the second term in the expansion, whose Feynman diagram 
is depicted in Fig.~1(c), contains an essential divergence of 
the $\omega$ meson self-energy type. 
This situation is the same as the vacuum polarization for
electron-positron field in QED except for propagating particle is
massive and the diagrams including not only vector mesons but also
scalar mesons contribute to the correction for baryon density.
According to Wichmann and Kroll\cite{WI56}, the one-loop vacuum
correction may be obtained by the sum of the finite part of
Fig.~1(c) and Fig.~1(a) subtracted by Fig.~1(c). Thus, the
renormalized baryon density from vacuum,
$\rho^{VP}_{\omega\hspace{0.5mm}{\rm ren}}(\vec{r})$, is written as
\begin{eqnarray}
\rho^{VP}_{\omega\hspace{1mm}{\rm ren}}(\vec{r})=
\rho^{(1)}_{\omega\hspace{1mm}{\rm ren}}(\vec{r})
+\rho^{(R)}_\omega(\vec{r}),
\label{finitew}
\end{eqnarray}
where $\rho^{(1)}_{\omega\hspace{1mm}{\rm ren}}(\vec{r})$, which
corresponds to the Uehling term\cite{UE35} in QED, denotes
the finite contribution from Fig.~1(c) and can be calculated
from the renormalized result in the momentum representation.
Choosing the renormalization point at $q^2 = 0$, 
only a wave function conterterm $\zeta_\omega$ is needed 
to obtain the finite result:
\begin{eqnarray}
\rho^{(1)}_{\omega\hspace{1mm}{\rm ren}}(\vec{r})=
\int \frac{d\vec{p}}{(2\pi)^3}e^{i\vec{p}\cdot\vec{r}}\omega_0(\vec{p})
\frac{g_\omega}{\pi^2}|\vec{p}|^2
\int_0^1 dx x(1-x) \ln
\left( 1+\frac{|\vec{p}|^2x(1-x)}{m_N^2}
\right),
\label{finitew1}
\end{eqnarray}
where $\omega_0(\vec{p})$ is the Fourier transform of
$\omega$ meson field $\omega_0(\vec{r})$. 
We estimate the contribution from the lowest order of $g_\omega^2$
by this explicit expression numerically.
The second term of Eq.~(\ref{finitew}), which
corresponds to the Wichmann-Kroll term\cite{WI56} in QED
is the residual finite density expressed by
\begin{eqnarray}
\rho^{(R)}_\omega(\vec{r})
&=&\rho^{VP}_\omega(\vec{r})-\rho^{(1)}_\omega(\vec{r})
\nonumber \\
&=& \int_{-i\infty}^{+i\infty}
\frac{dz}{2\pi i}{\rm Tr}[\gamma^0 G^H_V(\vec{r},\vec{r};z)]
%\nonumber
%\\
%&&
-  \int_{-i\infty}^{+i\infty}
\frac{dz}{2\pi i}{\rm Tr}[\int d\vec{x} \gamma^0 G^0(\vec{r},\vec{x};z)
g_\omega \gamma^0 \omega(\vec{x}) G^0(\vec{x},\vec{r};z)].
\label{finitewr}
\end{eqnarray}
In the present work, we evaluate the vacuum correction
for baryon density by using this expression directly.
After the partial-wave expansion by the Dirac angular-momentum quantum 
number $\kappa$, each $|\kappa|$ contribution of Eq.~(\ref{finitewr}) is still finite.
The partial-wave Green function of RHA calculated numerically 
on the imaginary axis is used in the first term, while 
the analytical form of the partial-wave Green function of 
the free Dirac equation is used in the second term.

There are several advantages to the method
solving the Green function on imaginary axis. For example,
we can carry out the $z$ integration without taking care of the poles.
In addition, we can employ Gaussian quadrature for the radial integral
in the second term of right-hand side of Eq.~(\ref{finitewr})
as well as $z$ integration
because the Green function on imaginary axis
behaves like the modified Bessel function which
is not an oscillating function.
As a result, the vacuum correction can be obtained very fast
with high precision
and it is possible to perform the RHA calculation with practical computational time.

It should be noted that $\rho^{(R)}_\omega(\vec{r})$ contain superficially
divergent contribution with the three $\omega$ mesons attached to the
baryon loop according to naive power counting,
which actually vanishes as a result of current conservation.
In QED, how to calculate this term has been discussed by many authors
\cite{WI56,GY72,RI75}
and it is well known that
this contribution vanishes if the summation over $\kappa$
is restricted to a finite number of terms.
Then, the $\kappa$ tail contribution is given by the extrapolation.
This conclusion is valid for the present case since the neutral $\omega$ meson
couples to the conserved baryon current.
Hence Eq.~(\ref{finitew}) together with (\ref{finitew1}) and 
(\ref{finitewr})
can be used to calculate the vacuum
correction for the baryon density numerically.

\subsection{Vacuum Correction for Scalar Density}

Next, we consider the vacuum correction for scalar density.
The regularization procedure of scalar density is performed under
the same concept as in the baryon density.
One can easily see that
the perturbative expansion of the Hartree vacuum correction for scalar density
corresponding to Eq.~(\ref{xpanw}) gives the Feynman diagrams depicted in Fig.~2,
where four diagrams from Fig.~2(b) to Fig.~2(e)
contain divergent contributions.

%\vspace{20mm}

The scalar density is renormalized by the counterterms $\alpha_1
\sigma(\vec{r})+1/2\alpha_2 \sigma^2(\vec{r}) +1/3\alpha_3
\sigma^3(\vec{r})+1/4\alpha_4 \sigma^4(\vec{r}) +1/2\zeta_\sigma
\partial_\mu\sigma(\vec{r})\partial^\mu\sigma(\vec{r})$ in $\delta{\cal L}$.
The finite part of Fig.~2(c) is calculated  
from the renormalized result in the vacuum.
The mass ($\alpha_2$) and
wave function ($\zeta_\sigma$) counterterms are required to make 
the contribution finite.   
The result is given by
\begin{eqnarray}
\rho^{(1)}_{\sigma\hspace{1mm}{\rm ren}}(\vec{r})=
\int \frac{d\vec{p}}{(2\pi)^3}e^{i\vec{p}\cdot\vec{r}}\sigma(\vec{p})
\frac{3g^2_\sigma}{2\pi^2}
\left[
\frac{1}{6}|\vec{p}|^2-
\int_0^1 dx (m_N^2+|\vec{p}|^2x(1-x)) \ln
\left( 1+\frac{|\vec{p}|^2x(1-x)}{m_N^2}
\right)
\right],
\label{finites1}
\end{eqnarray}
where $\sigma(\vec{p})$ is the Fourier transform of $\sigma(\vec{r})$
and we have chosen $q^2 = 0$ for the renormalization point.
The physical vacuum contribution for scalar density is given 
by the sum of 
$\rho^{(1)}_{\sigma\hspace{1mm}{\rm ren}}(\vec{r})$ and the residual
finite density denoted by $\rho^{(R)}_\sigma(\vec{r})$:
\begin{eqnarray}
\rho^{VP}_{\sigma\hspace{1mm}{\rm ren}}(\vec{r})=
\rho^{(1)}_{\sigma\hspace{1mm}{\rm ren}}(\vec{r})
+\rho^{(R)}_\sigma(\vec{r}).
\label{finites}
\end{eqnarray}
where the residual finite density $\rho^{(R)}_\sigma(\vec{r})$ 
includes the finite contributions arising from Figs.~2(b), 2(d), and
2(e).
The finite residual vacuum density
$\rho^{(R)}_\sigma(\vec{r})$ is evaluated by subtracting the 
contribution of Fig.~2(c) and the counterterms
from the unrenormalized divergent scalar density: 
\begin{eqnarray}
\rho^{(R)}_\sigma(\vec{r})
&=&
\int_{-i\infty}^{+i\infty}
\frac{dz}{2\pi i}{\rm Tr}[G^H_V(\vec{r},\vec{r};z)]
\nonumber \\
&& -  \int_{-i\infty}^{+i\infty}
\frac{dz}{2\pi i}{\rm Tr}[\int d\vec{x} G^0(\vec{r},\vec{x};z)
g_\sigma \sigma(\vec{x}) G^0(\vec{x},\vec{r};z)]
\nonumber \\
&& 
-(\alpha_1+\alpha_3\sigma^2(\vec{r})+\alpha_4\sigma^3(\vec{r}))/g_\sigma,
\label{finitesr} 
\end{eqnarray}
where the coefficients $\alpha_1, \alpha_3$, and $\alpha_4$ of the counterterms are given by 
\begin{eqnarray}
\alpha_1
&=&
g_\sigma \int_{-i\infty}^{+i\infty}
\frac{dz}{2\pi i}{\rm Tr}[G^0(\vec{r},\vec{r};z)]
%\nonumber \\
\label{alpha1} \\
\alpha_3
&=&
g_\sigma^3   \int_{-i\infty}^{+i\infty}
\frac{dz}{2\pi i}{\rm Tr}[\int d\vec{x_1}d\vec{x_2}  G^0(\vec{r},\vec{x_1};z)
G^0(\vec{x_1},\vec{x_2};z)
G^0(\vec{x_2},\vec{r};z)]
\label{alpha3} \\
\alpha_4
&=&
g_\sigma^4  \int_{-i\infty}^{+i\infty}
\frac{dz}{2\pi i}{\rm Tr}[\int d\vec{x_1}d\vec{x_2}d\vec{x_3}G^0(\vec{r},\vec{x_1};z)
G^0(\vec{x_1},\vec{x_2};z)
G^0(\vec{x_2},\vec{x_3};z)
G^0(\vec{x_3},\vec{r};z)].
\label{alpha4}
\end{eqnarray}
We note that $\alpha_1$, $\alpha_3$, and $\alpha_4$ are
independent of $\vec{r}$ though it remains in the right-hand side
of these expressions. Performing the partial-wave expansion of
free Green functions, however, each $|\kappa|$ contribution to
the coefficients depends on the radial part $r$. With
these partial-wave subtraction terms, 
%{\it constants}, 
the scalar
density is calculated for each $|\kappa|$. The net effect of
scalar density generated from the vacuum polarization are obtained
by taking the extrapolation $|\kappa_{max}|\rightarrow\infty$.

\section{Computational detail for vacuum correction}

As has been explained in the previous section, 
$\rho^{(1)}_{\omega\hspace{1mm}{\rm ren}}$ and
$\rho^{(1)}_{\sigma\hspace{1mm}{\rm ren}}$ contributions are  
calculated from the explicit forms (\ref{finitew1}) and
(\ref{finites1}). They contribute by a small amount as discussed
below.
Therefore, here we explain mainly the
estimation of the residual contributions defined by
Eqs.~(\ref{finitewr}) and (\ref{finitesr}). 
We write the radial part of the residual baryon and scalar densities as
\begin{eqnarray}
\rho^{R}_{\omega\hspace{1mm}{\rm ren}}({r})=
\sum_{|\kappa|=1}^{|\kappa_{max}|}
\rho^{R}_{\omega\hspace{0.5mm}{\rm ren},|\kappa|}({r})
\end{eqnarray}
and
\begin{eqnarray}
\rho^{R}_{\sigma\hspace{1mm}{\rm ren}}({r})=
\sum_{|\kappa|=1}^{|\kappa_{max}|}
\rho^{R}_{\sigma\hspace{0.5mm}{\rm ren},|\kappa|}({r}),
\end{eqnarray}
respectively. Here, $\rho^{R}_{\omega\hspace{0.5mm}{\rm
ren},|\kappa|}$ ($\rho^{R}_{\sigma\hspace{0.5mm}{\rm
ren},|\kappa|}$) represents the contribution from
$|\kappa|=\pm \kappa$ to net residual baryon (scalar) density. We
compute $\rho^{R}_{\omega\hspace{0.5mm}{\rm ren},|\kappa|}$ and
$\rho^{R}_{\sigma\hspace{0.5mm}{\rm ren},|\kappa|}$ using the
angular momentum decomposed expression of Eqs.~(\ref{finitewr})
and (\ref{finitesr}). The detail of our calculation is the
following. For the respective contributions of $|\kappa|$ in
Eqs.~(\ref{finitewr}) and (\ref{finitesr}), we carry out the
integral over $iz$ using Gaussian quadrature. 
The radial Green functions with the imaginary energy are 
given in terms of the two solutions of
the Dirac equation; the solution regular at $r=0$ and the solution
regular
at $r=\infty$. Several values of the upper and lower limits of the
integral over $z$ are selected depending on the radius parameter
$r$ around $20i \sim 50i$ GeV. 
The vacuum densities $\rho^{R}_{\omega\hspace{0.5mm}{\rm
ren},|\kappa|}({r})$ and $\rho^{R}_{\sigma\hspace{0.5mm}{\rm
ren},|\kappa|}({r})$ for $z_{max}\rightarrow \infty$ 
are extrapolated from the resulting integrated values.
Then, the sum over
$|\kappa|$ is performed until the cutoff $|\kappa_{max}|$. 
The sum over
$|\kappa|$ does not converge so rapidly. In the
present work, $|\kappa_{max}|$ in baryon density is 33
for both of $^{16}$O and $^{40}$Ca, while $|\kappa_{max}|$
in scalar density is 36 for $^{16}$O and 48 for
$^{40}$Ca.  Finally,
we have extrapolated $|\kappa|$ contribution to larger values
to obtain the convergent vacuum densities.
As an example, the contributions from
$|\kappa|=$5, 10, 15, 20, 25, and 30 for  in $^{16}$O are shown in
Fig.~3.

%\vspace{30mm}

An accuracy test for the computed vacuum correction of the baryon density
$\rho^{R}_{\omega\hspace{0.5mm}{\rm ren}}({r})$
is provided by the requirement that for the total
induced vacuum correction should vanish due to the conservation
of baryon density,
\begin{eqnarray}
\Delta B=4\pi\int_0^\infty dr r^2
\rho^{R}_{\omega\hspace{0.5mm}{\rm ren}}({r})=0.
\end{eqnarray}
In the present calculation with various QHD parameters,
we found $\Delta B \sim 10^{-3}$ as a typical values both for $^{16}$O and $^{40}$Ca.
For the scalar density, on the other hand, we have no requirement 
of conservation laws.
However, it would be reasonable to expect
the numerical error in scalar density is of the same order of magnitude
as the one in the baryon density.

In Sec.~II.B, we mentioned that the divergence of unrenormalized
baryon density in the present model has the same structure as
unrenormalized vacuum charge density of electron-positron field in
the QED correction. However, the dependence on the partial-wave
contribution is largely different from the QED case: a large
$|\kappa_{max}|$ is required to achieve the convergence as seen in
Fig.~3, while for the renormalized charge density in QED only the
term with $|\kappa_{max}|=1$ is a good approximation\cite{GY72}.
In addition, it should be pointed out that Eqs.~(\ref{finitew1}) and
(\ref{finites1}) are negligible in the present calculation,
namely, $\rho^{(1)}_{\omega\hspace{1mm}{\rm ren}}$ and
$\rho^{(1)}_{\sigma\hspace{1mm}{\rm ren}}$ are an order of
magnitude smaller than the residual densities in contrast with the
QED calculation, in which the corresponding contribution known as
the Uehling effect has a dominant contribution\cite{RI75}. These
differences from QED seem to be caused by the $\omega-\sigma$
couplings as well as the large coupling constant of the
nuclear force. In particular, the baryon
density, the source of the $\omega$ meson, is strongly influenced
by the $\sigma$ meson. 
This is shown in Fig.~4 where
the baryon densities $\rho^{(1)}_{\omega\hspace{1mm}{\rm ren}}(r)$ 
with and without $\sigma$
meson are compared. We can see there that the vacuum
correction for baryon density is negligibly small if the $\omega$
meson only is taken into account. We found that the baryon density
induced by the vacuum polarization comes from the diagrams
with $\sigma$ meson self-energy insertions mainly.

%\vspace{10mm}

\section{Results and Discussion }

Here we show the results of the relativistic Hartree calculation
with a rigorous treatment of renormalized densities in $^{16}$O and
$^{40}$Ca based on the Lagrangian density of Eq.~(\ref{nucleusL}). 
The numerical procedure of the present RHA calculation is similar 
to that used in the conventional RMF calculation:
first, the Dirac equation (\ref{Dirac}) for only the valence nucleons 
is solved under the external $\omega$ and $\sigma$ fields. 
Second, using the same potential, we calculate vacuum densities 
by means of the Green function described in the previous section. 
Third, the equations of motion of mesons, (\ref{KGw}) and (\ref{KGs}), 
are solved with the source terms due to the valence nucleons and vacuum
contributions. Substituting these results in the Dirac equation, 
we complete one iteration step.
Using the same method as indicated in the previous section, 
the energy shift due to the vacuum polarization is estimated by
\begin{eqnarray}
E_{VP}&=&\int d\vec{r}
\left[
\int_{-i\infty}^{+i\infty}\frac{dz}{2\pi i}{\rm Tr}[\gamma^0
\left(G^H_V(\vec{r},\vec{r};z)-G^0(\vec{r},\vec{r};z)\right)]z
+(CT)
\right],\\
&&CT=
-1/4\zeta_\omega(\vec{\nabla}\omega_{0})^2
+1/2\zeta_\sigma(\vec{\nabla}\sigma)^2
+\alpha_1\sigma
+1/2\alpha_2\sigma^2
+1/3\alpha_3\sigma^3
+1/4\alpha_4\sigma^4,
\label{EVP}
\end{eqnarray}
is calculated at each step of iteration.
The iteration is continued until the
total binding energy of the nucleus  
$E_{\rm{total}}  = E_{\rm{meson}}  + E_{\rm{valence}} + E_{VP}$
converges showing self consistency.

The QHD parameter set is chosen so as to reproduce reasonably well the
experimental values of the total binding energies, the rms
radii, and the single-particle energies for both of $^{16}$O and
$^{40}$Ca. In the second column of Table I, we
give the results with the coupling constants and masses
$g_\sigma=7.38$, $g_2=7.90$, $g_3=3.20$, and $m_\sigma=458.0$ MeV
for $\sigma$ meson, and $g_\omega=9.18$ and $m_\omega=783.0$ MeV
for $\omega$ meson. The results of the RMF calculation with the
parameter set TM2\cite{SU94} and experimental data taken from
Ref.~\cite{DATA} are also shown in the third and last columns,
respectively. We see that our total binding energies including
the vacuum correction and rms radii are similar to those of RMF and
agree with the experimental data well. 
However, it should be noted that the
present result is produced by using smaller coupling constants
than those of RMF. 
Hence, the scalar and vector potentials in the present
RHA differ considerably from those of RMF. 
The results of the present RHA in $^{16}$O
and $^{40}$Ca are compared with the results of
RMF in Figs.~5 and 6 for scalar and
vector potentials, respectively.
One can see that the fields in the RHA are
considerably small compared with those in the RMF.

%\vspace{25mm}

The vacuum contribution plays a crucial role for the generation of
the weak meson fields. In order to see this, we plot
$\sigma$-meson field in nuclear matter as a function of the
coupling constant $g_\sigma$ while keeping the other parameters
fixed in Fig.~7. The figure shows that 
the vacuum correction contributes destructively
to the valence contribution and 
it is impossible to obtain a strong meson field
unless a very large coupling constant is used. 
Of course, we could choose a
parameter set with large coupling constants. However, such a
parameter set would have produced an unstable solution in the RHA
calculation, because if the deeply-bound
antinucleon states are produced by the strong $\omega$ and $\sigma$ 
fields 
with large coupling constants, then the vacuum effect, 
which works in the opposite direction, 
also becomes large to suppress the strong fields. Such a RHA solution is not realistic because it is
unstable even for the trivial fluctuation of the density. Thus, we
have to choose a parameter set which produces a weak field in the 
self consistent iteration.

%\vspace{20mm}

The RMF reproduces reasonably well the observed tendency of the 
single-particle spectra due to the small effective mass of the nucleon,
$m^*_N(r)=m_N+g_\sigma\sigma (r)$. 
The fact that the scalar field
is suppressed in RHA results in the large effective mass. As a result, 
it raises a problem in fitting the single-particle energies. As seen
in Table I, the energy splittings in the single-particle states of the
present RHA are very small and they are unlikely to agree with the
experimental values. This problem has been already known by RHA
calculation using the local-density approximation 
and the derivative expansion to estimate the vacuum correction\cite{WA88,MA99}. 
It is difficult
to resolve this problem by RHA with ordinary QHD models used up to now.

Thus, the QHD models require a mechanism for the spin-orbit splittings
other than the small effective mass. 
One suggestion is made in
Refs.~\cite{MA03,SU942}, where a tensor-coupling of $\omega$ meson
is introduced in order to provide the spin-orbit splittings.
However, the omega meson coupling with the nucleon is known to be
dominated by the vector coupling in the nucleon-nucleon potential.
Another candidate to solve this problem may be the possibility of
the finite pion mean field in the relativistic Hartree framework,
which was suggested to provide the spin up and spin down partners
to have large energy separations\cite{OG04}. It is interesting to
extend the present RHA calculation with these effects taken into
account. This is certainly a subject to be worked
out in future study.

\section{Comparison with the previous method for vacuum polarization}

The effect of the
negative-energy nucleons for finite nuclei 
was first estimated by the local-density
approximation\cite{HO84,IC88,FO89}. It was developed further by
applying the derivative-expansion method \cite{PE86,WA88,MA99,MA03}. 
In this section, we compare our results densities 
induced by the vacuum polarization with those of the local-density approximation
and the derivative-expansion method. The local-density
approximation uses as input the result of the infinite nuclear
matter. In this approximation, the vacuum correction 
is given by
\begin{eqnarray}
\rho_{\sigma,ren}^{VP(LDA)}(r) = -\frac{1}{\pi^2}
\left[m^{*3}_N\ln(m^*_N / m_N)+1/3m_N^3-3/2m_N^2 m^*_N+3m_N m^{*2}_N
-11/6 m^{*3}_N\right],
\label{LDA}
\end{eqnarray}
and decreases the
scalar density in nuclear interior. 
The vacuum does not change the baryon density because of the conservation
of the baryon number. In the derivative-expansion method, on the other hand,
the presence of the derivative term allows non-vanishing correction for baryon density
as well as for scalar density:
\begin{eqnarray}
\rho_{\omega,ren}^{VP(DE)}(r) &=& -\frac{g_\omega}{3\pi^2}
\vec{\nabla}\cdot\ln\left(\frac{m_N^*(r)}{m_N}\right)\vec{\nabla}\omega_0(r),
\label{DEw}
\\
\rho_{\sigma,ren}^{VP(DE)}(r) &=&
\rho_{\sigma,ren}^{VP(LDA)}(r) -\frac{g_\sigma}{2\pi^2}
\vec{\nabla}\cdot\ln\left(\frac{m_N^*(r)}{m_N}\right)\vec{\nabla}\sigma(r)
\nonumber \\
&&
-\frac{g_\sigma^2}{4\pi^2 m_N^*(r)}
(\vec{\nabla}\sigma(r))^2
+\frac{g_\omega^2}{6\pi^2 m_N^*(r)}
(\vec{\nabla}\omega(r))^2,
\label{DEs}
\end{eqnarray}
where the leading order of the derivative terms is taken into
account. The baryon and scalar densities induced by the
vacuum polarization are given in Fig.~8, together with those from
the local-density approximation and the derivative expansion. For
purpose of comparison, we assume there the same potential in
evaluating the vacuum polarization. We can see that the densities
obtained by the local-density approximation are corrected significantly
not only for baryon density, which vanishes in this approximation,
but also for scalar density. 
The both of scalar and baryon density-profiles obtained 
by the present calculation are in a surprisingly good agreement
with those of the derivative expansion.

However, this cannot be always the case\cite{ST97}
and it is possible to attribute the excellent agreement 
between our method and the leading-order derivative expansion
to the character of $\sigma-\omega$ model.
Consider, for example, the vacuum correction in the baryon density
without $\sigma$ meson.
In this situation, the vacuum correction using the present method 
can be significant with large coupling constant of $\omega$ meson.
As found in Eq.~(\ref{DEw}), on the other hand,
the vacuum correction from the derivative expansion vanishes exactly
for $m_N^*(r)\rightarrow m_N$.
Hence, we find that $\sigma$ meson plays important role
in the agreement between our method and the leading-order derivative expansion.
%In the present RHA model, a net potential is produced by the scalar
%and vector potentials with opposite sign.
%This fact has consequently lead to our result in the proceeding section that the
%diagrams connected with $\sigma$ meson are dominant in baryon density.
Thus, the present calculation supports that leading-order derivative expansion
is greatly useful for the estimation of the vacuum correction in RHA.

In the present calculation of the full RHA for finite nuclei, 
the vacuum-polarization corrections (\ref{finitew1}) and (\ref{finites1}) 
to the meson propagators are implicitly taken into account to 
all orders through iteration to achieve the self-consistency in the relativistic
Hartree approximation.
There the unphysical pole in the meson propagator at finite momentum transfer, 
known as Landau ghost, may affect the present numerical
results\cite{IC88,KT91} through integration over the momentum transfer.
However, this unphysical effect is not significant in the RHA of
finite nuclei, because even if the (\ref{finitew1}) and (\ref{finites1}) 
terms are totally ignored in the 
calculation, the final results of scalar and vector densities do not change appreciably.
The good agreement with the results of the derivative expansion, 
where the Landau ghost plays no role, also implies that the
unphysical effect is negligible.

%\vspace{20mm}

\section{Summary}

We have developed a rigorous method to calculate 
vacuum-polarization effects in relativistic Hartree approach.
The renormalized baryon and scalar densities have been 
evaluated within a practical computational time 
replacing the summation of the Hartree basis by the numerical integral 
of the Dirac Green function over the imaginary energy.
We have obtained numerical results that the vacuum corrections for
baryon and scalar densities are non-negligible in the RHA
calculation. In particular, we have seen that the vector and
scalar potentials in the RHA are largely suppressed by the
feedback of the vacuum polarization to the mean fields. Such
suppressed potentials of the $\omega$ and $\sigma$ fields affect
the theoretical calculations for binding energies of antinucleon
and the various sum rules.

Our results with the Walecka model have reproduced the
experimental binding energies and rms radii of $^{16}$O and
$^{40}$Ca nicely after adjusting the parameters.
However, it was impossible to find a QHD parameter set to reproduce
spin-orbit splittings in accordance with the observed data
and required by nuclear shell model.
In the $\sigma-\omega$ model, the main attraction is caused 
by the large $\sigma$ mean field, which provides small nucleon 
effective mass in the finite nuclei.
However, the negative-energy nucleons do not like to change its
mass from the free value.  Hence, the negative-energy nucleons try
to keep the nucleon mass at the free value. For the net effect,
the effective nucleon mass remains to be quite large. Once the
effective nucleon mass becomes large, the spin-orbit splittings in
the single particle spectra come out to be very small.
The QHD type effective theory based on the $\sigma-\omega$ mesons
need to include new type of interaction terms and/or to go beyond the RHA
approximation to solve this problem.

We have found that our result of RHA calculation is very similar 
to that in Refs.~\cite{PE86,MA99} where the derivative-expansion 
method was used to estimate the vacuum polarization. 
In particular, it has been shown that the agreement of density-profiles 
of the vacuum correction is quite good. 
Thus, the validity of this approximation has been confirmed
by the present calculation.

\vskip30pt
\
\

%\begin{acknowledgments}

This work has been supported by MATSUO FOUNDATION, Suginami, Tokyo.
A.H. would like to thank Prof. S. Kita and Prof. Y. Tanaka for giving him the opportunity
to do this work and useful discussion. Y.H. acknowledges Dr. K. Tanaka for
useful discussion.

%\end{acknowledgments}

%\vfill\eject

\clearpage \setlength{\baselineskip}{0.0in}

TABLE I.  The total binding energies, the rms charge radii, and
the single-particle energies in $^{16}$O and $^{40}$Ca.
\vspace{1mm}\\
\begin{math}
\begin{array}{lcccc}
\hline\hline & \hspace{10mm}{\rm Present\hspace{1mm}RHA}
\hspace{10mm} & \hspace{10mm}{\rm TM2\hspace{1mm}}\hspace{10mm}
& \hspace{10mm}{\rm Experiment}\hspace{10mm}&\\
\hline
^{16}{\rm O}\\
E_{\rm total}/A (E_{VP}/A) \hspace{1mm}{\rm [MeV]} & 8.05 (1.69)
& 7.93 ( - )
& 7.98 ( - )\\
r_{ch}\hspace{1mm}{\rm [fm]}& 2.65 & 2.67
& 2.74\\
{\rm Single\hspace{1mm}particle
\hspace{1mm}state\hspace{1mm}of\hspace{1mm}proton}\\
1s_{1/2}{\rm [MeV]} & 31.0 & 38.2
& 40\pm 8\\
1p_{3/2}{\rm [MeV]} & 15.6 & 18.6
& 18.4\\
1p_{1/2}{\rm [MeV]} & 13.3 & 11.1
& 12.1\\
{\rm Single\hspace{1mm}particle
\hspace{1mm}state\hspace{1mm}of\hspace{1mm}neutron}\\
1s_{1/2}{\rm [MeV]} & 35.6 & 42.3
& 45.7\\
1p_{3/2}{\rm [MeV]} & 19.7 & 22.4
& 21.8\\
1p_{1/2}{\rm [MeV]} & 17.4 & 14.8
& 15.7\\
\hline
^{40}{\rm Ca}\\
E_{\rm total}/A (E_{VP}/A) \hspace{1mm}{\rm [MeV]} & 8.47 (2.23)
& 8.48 ( - )
& 8.55 ( - )\\
r_{ch}\hspace{1mm}{\rm [fm]}& 3.42 & 3.50
& 3.45\\
{\rm Single\hspace{1mm}particle
\hspace{1mm}state\hspace{1mm}of\hspace{1mm}proton}\\
1s_{1/2}{\rm [MeV]} & 36.5 & 45.2
&50\pm 11\\
1p_{3/2}{\rm [MeV]} & 25.5 & 30.7
&\\
1p_{1/2}{\rm [MeV]} & 24.0 & 36.2
&34\pm 6\\
1d_{5/2}{\rm [MeV]} & 13.5 & 16.1
&\\
1d_{3/2}{\rm [MeV]} & 11.0 & 8.7
&8.3\\
2s_{1/2}{\rm [MeV]} & 9.1 & 8.5
&10.9\\
{\rm Single\hspace{1mm}particle
\hspace{1mm}state\hspace{1mm}of\hspace{1mm}neutron}\\
1s_{1/2}{\rm [MeV]} & 45.5 & 53.1
&\\
1p_{3/2}{\rm [MeV]} & 33.8 & 38.3
&\\
1p_{1/2}{\rm [MeV]} & 32.3 & 33.8
&\\
1d_{5/2}{\rm [MeV]} & 21.2 & 23.4
&\\
1d_{3/2}{\rm [MeV]} & 18.7 & 15.9
&15.6\\
2s_{1/2}{\rm [MeV]} & 16.9 & 15.6
&18.1\\
\hline\hline\\
\end{array}
\end{math}

\clearpage

\begin{figure}[h]
\centering
\includegraphics[width=9.0cm,clip]{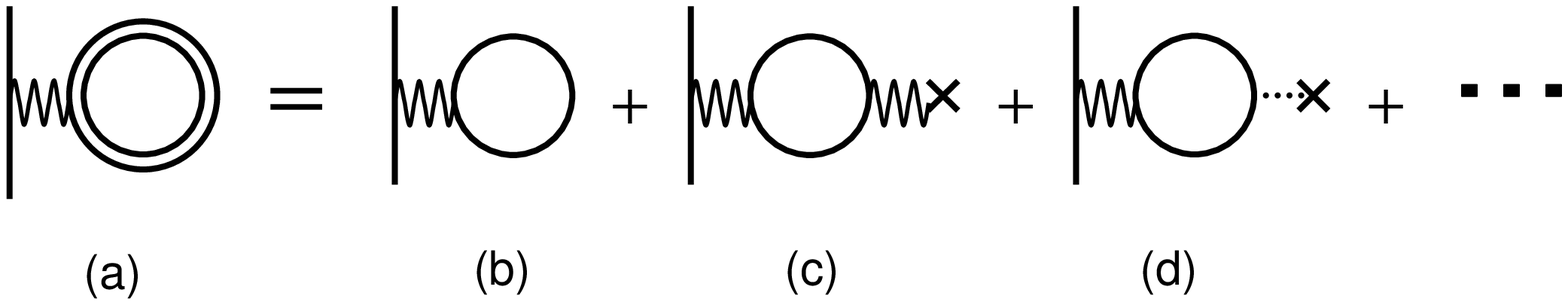}
\caption{\label{fig1}
Graphical representation of
the baryon density in the self-consistent relativistic
Hartree approximation.
The double and single lines denote the Hartree propagator 
and the free propagator of nucleon, respectively.
The wavy and dotted lines with cross represnt 
the vector and scalar potentials given by the previous step of
Hartree iteration respectively. 
The divergence contained in (a) is caused by the contribution from (c).}
\end{figure}

\clearpage

\begin{figure}[h]
\centering
\includegraphics[width=9.0cm,clip]{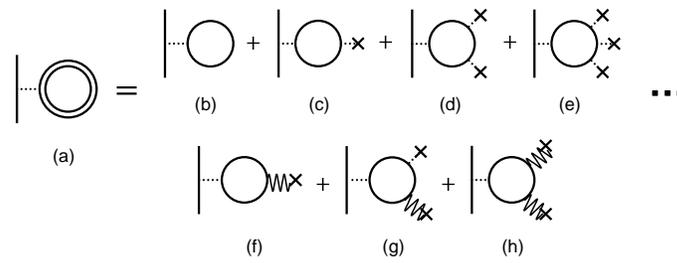}
\caption{\label{fig2}
Graphical representation of
the scalar density in the self-consistent relativistic
Hartree approximation.
Same notations as in Fig.~1 are used.
The divergence contained in (a) is caused by
the contributions from (b) to (e).}
\end{figure}

\clearpage

\begin{figure}[h]
\centering
\includegraphics[width=6.0cm,clip]{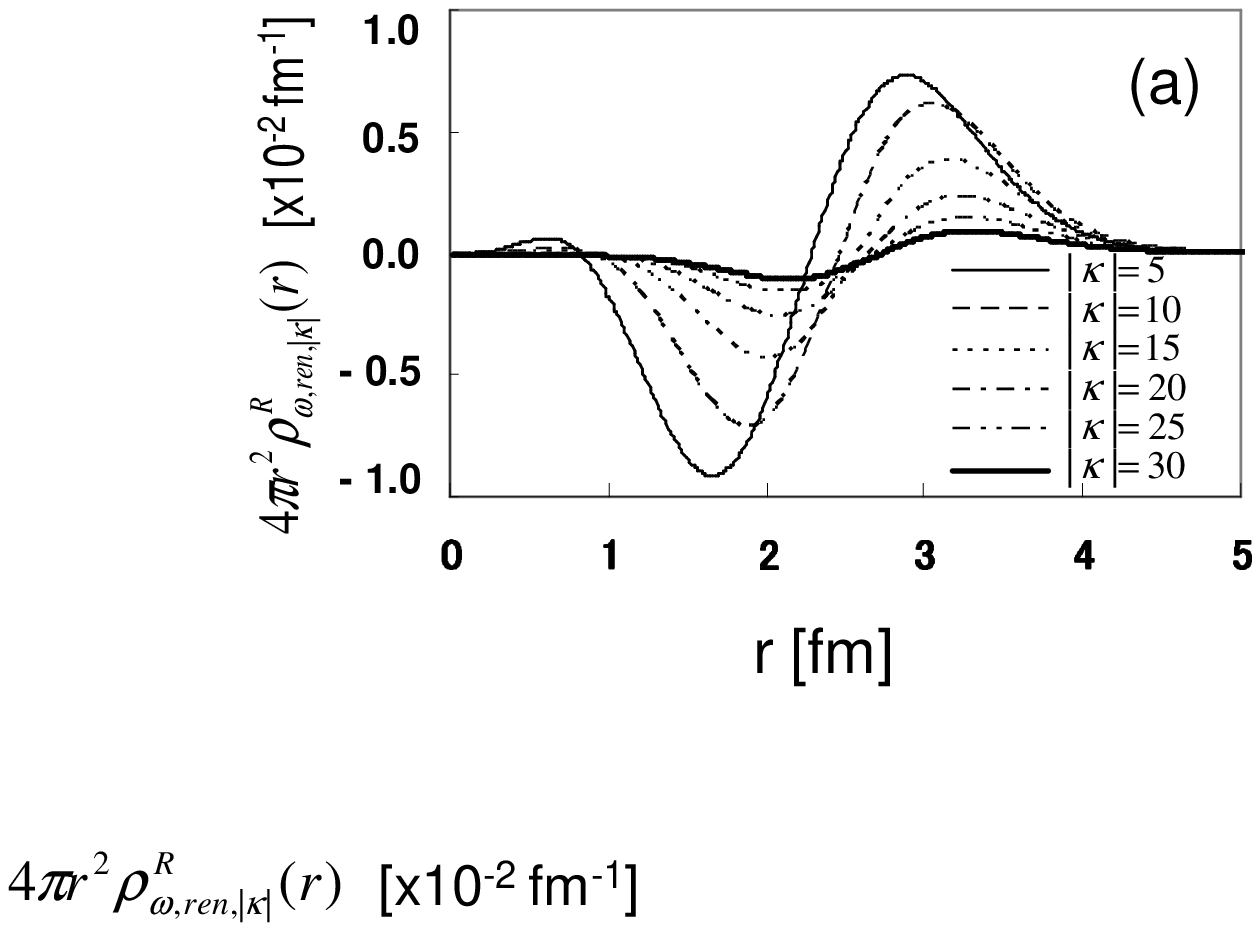}
\end{figure}

\vspace{20mm}
\begin{figure}[h]
\centering
\includegraphics[width=6.0cm,clip]{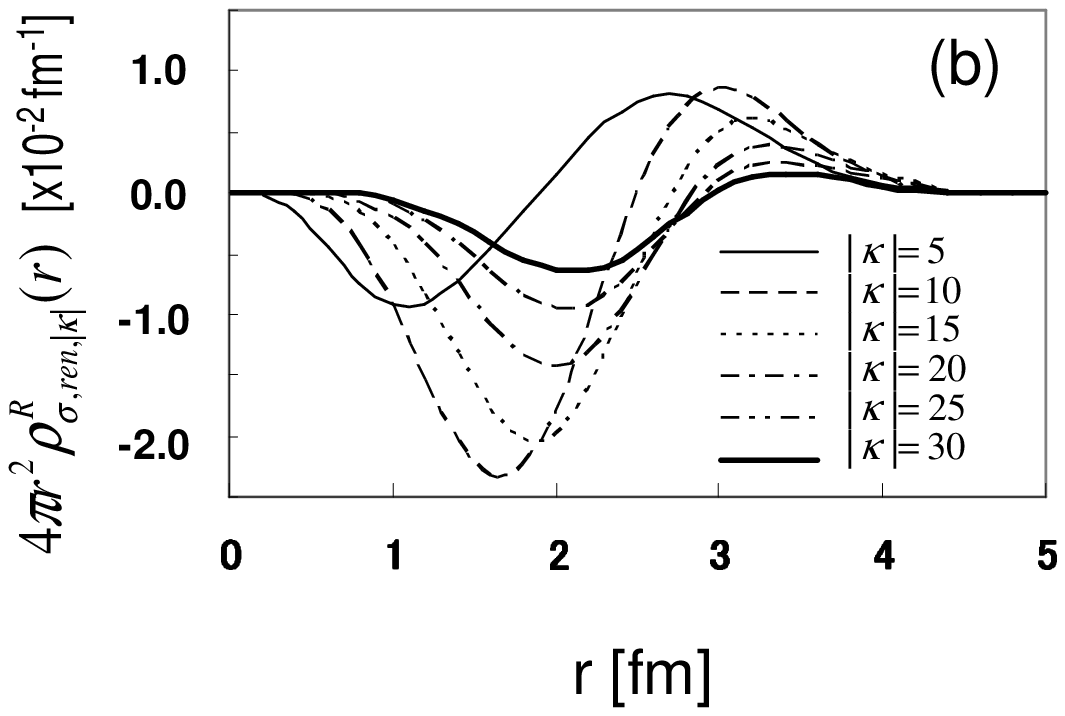}
\caption{\label{fig3}
The contributions from $|\kappa|=$5, 10, 15, 20, 25, and 30
to (a) the baryon density 
$\rho^{R}_{\omega\hspace{0.5mm}{\rm ren},|\kappa|}({r})$
and (b) the scalar density 
$\rho^{R}_{\sigma\hspace{0.5mm}{\rm ren},|\kappa|}({r})$.}
\end{figure}

\clearpage

\begin{figure}[h]
\centering
\includegraphics[width=10.0cm,clip]{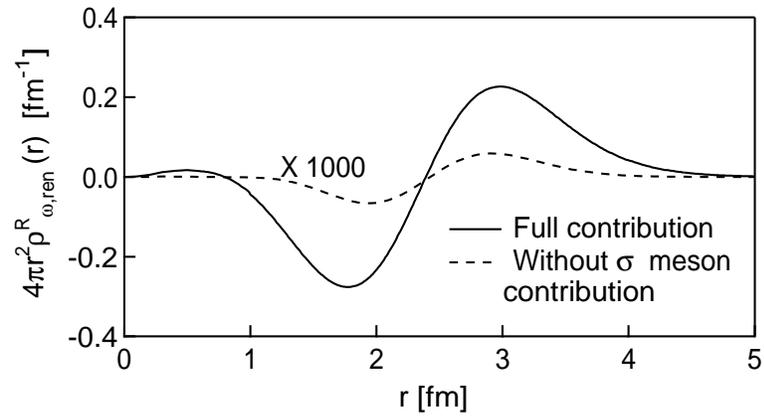}
\caption{\label{fig4}
Comparison between
the baryon densities with
and without $\sigma$ meson contribution.
The latter is given by solving the Dirac Green function
with the $\omega$ meson field only.}
\end{figure}

\clearpage

\begin{figure}[h]
\centering
\includegraphics[width=7.0cm,clip]{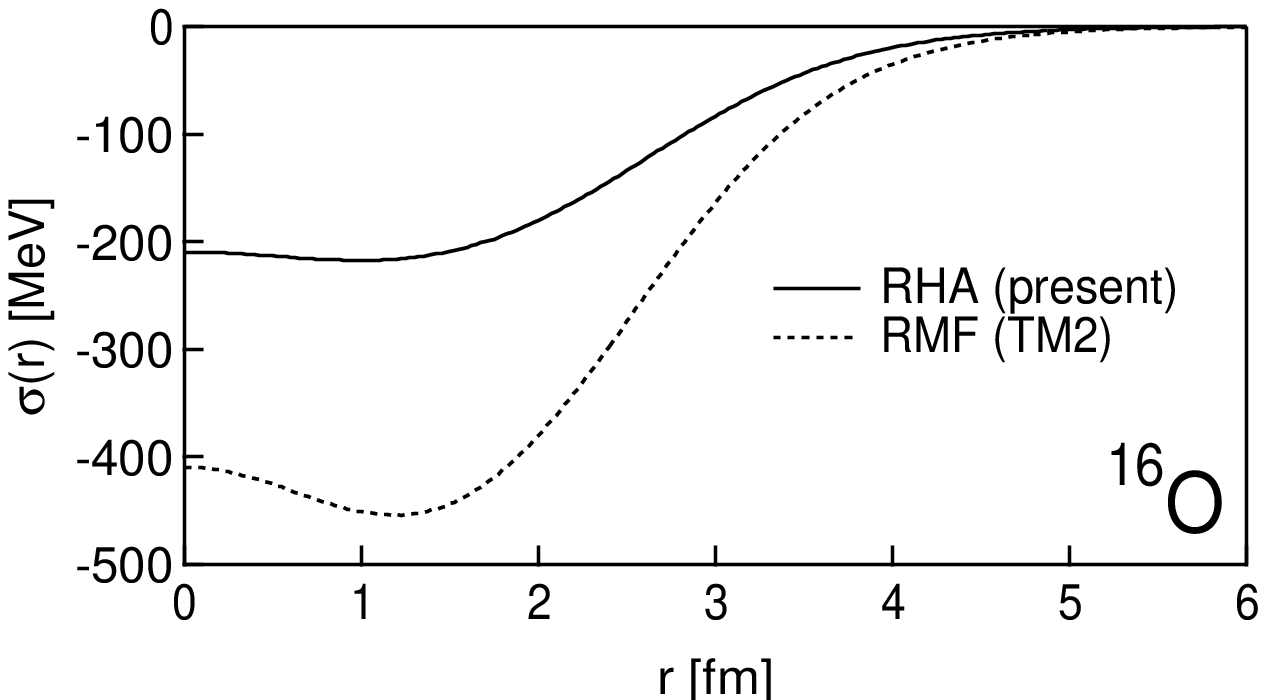}
\includegraphics[width=7.0cm,clip]{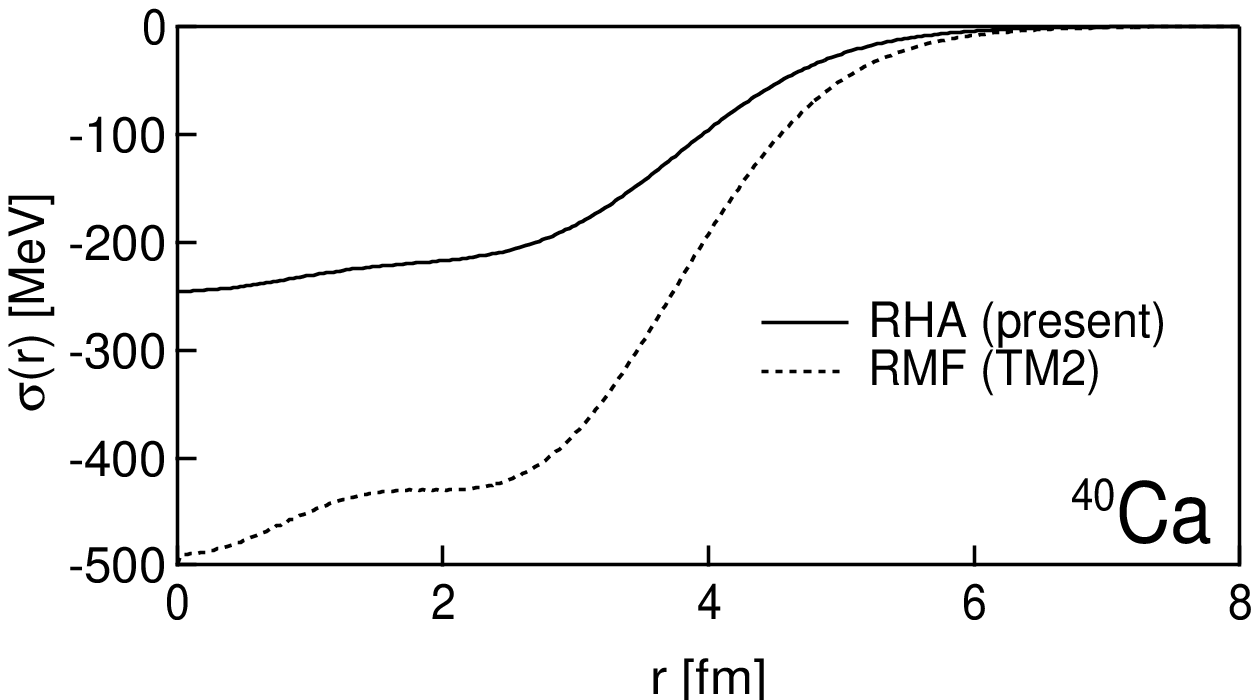}
\caption{\label{fig5} The scalar potentials for $^{16}$O in the
left panel and for $^{40}$Ca in the right panel.  The scalar
potential with the vacuum polarization is shown by solid curve,
while the one without it by dashed curve.}
\end{figure}

\clearpage

\begin{figure}[h]
\centering
\includegraphics[width=7.0cm,clip]{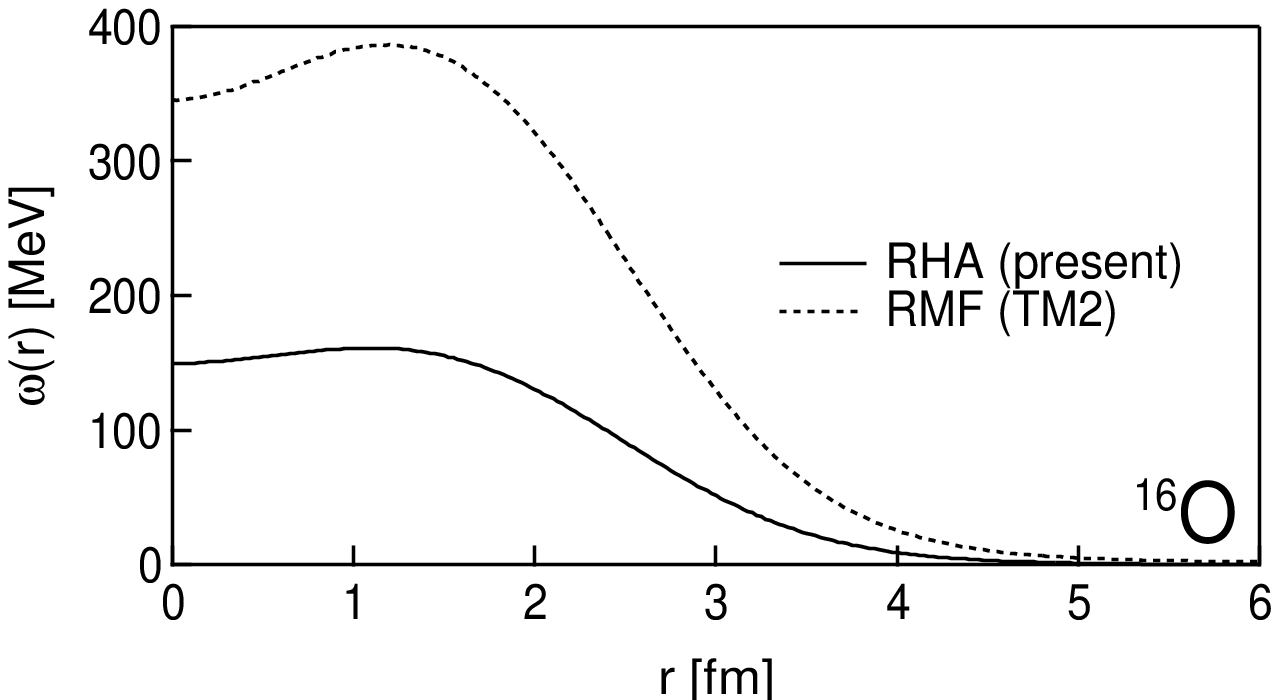}
\includegraphics[width=7.0cm,clip]{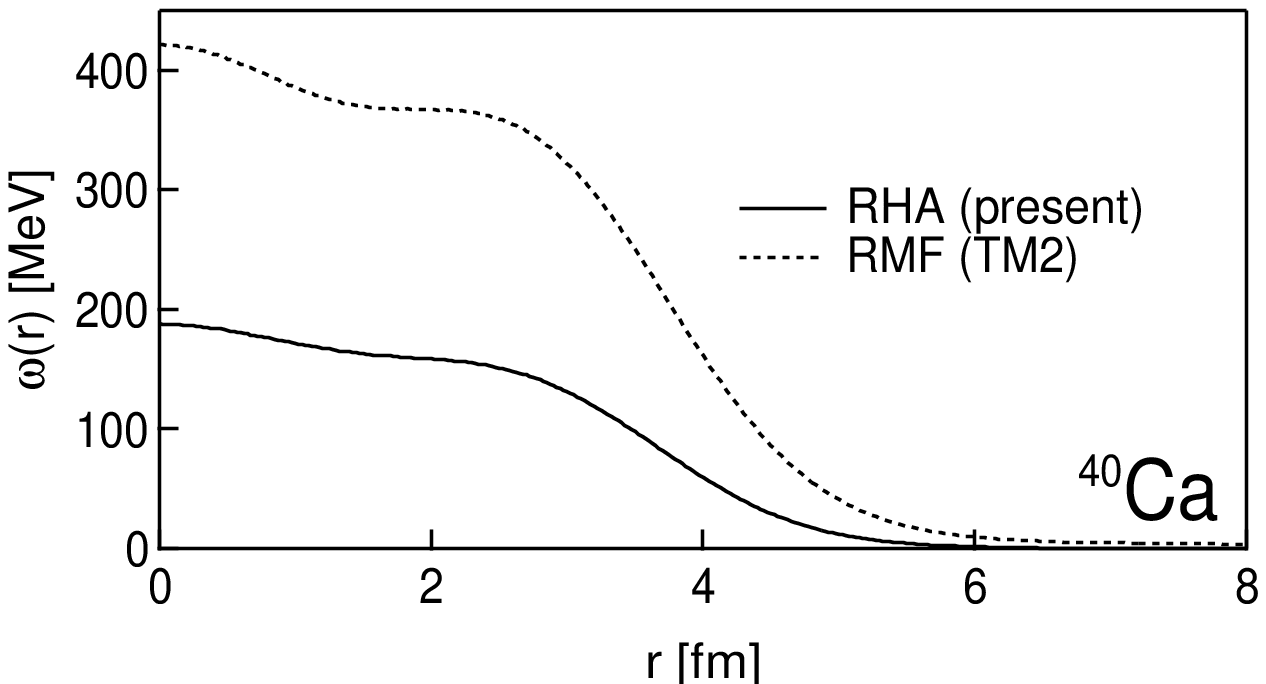}
\caption{\label{fig6} The vector potentials for $^{16}$O in the
left panel and for $^{40}$Ca in the right panel.  The scalar
potential with the vacuum polarization is shown by solid curve,
while the one without it by dashed curve.}
\end{figure}

\clearpage

\begin{figure}[h]
\centering
\includegraphics[width=7.0cm,clip]{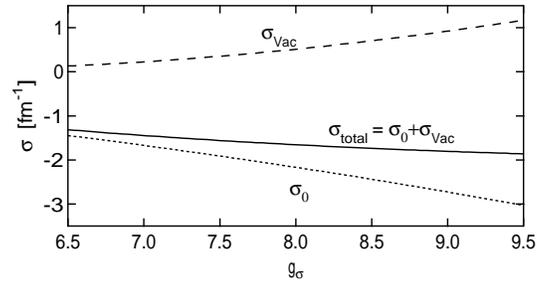}
\caption{\label{fig7} The scalar potential in nuclear matter.
$\sigma$ meson mass $m_\sigma=458.0$ and Fermi momentum $k_F=1.42$
are employed. $\sigma_0$ denotes the ordinary $\sigma$ meson field
generated from the valence nucleons while $\sigma_{Vac}$ denotes
the contribution from the vacuum. Due to the cancellation between
them, the net $\sigma$ meson field does not increase smoothly with
the coupling constant, $g_\sigma$.}
\end{figure}

\clearpage

\begin{figure}[h]
\centering
\includegraphics[width=7.0cm,clip]{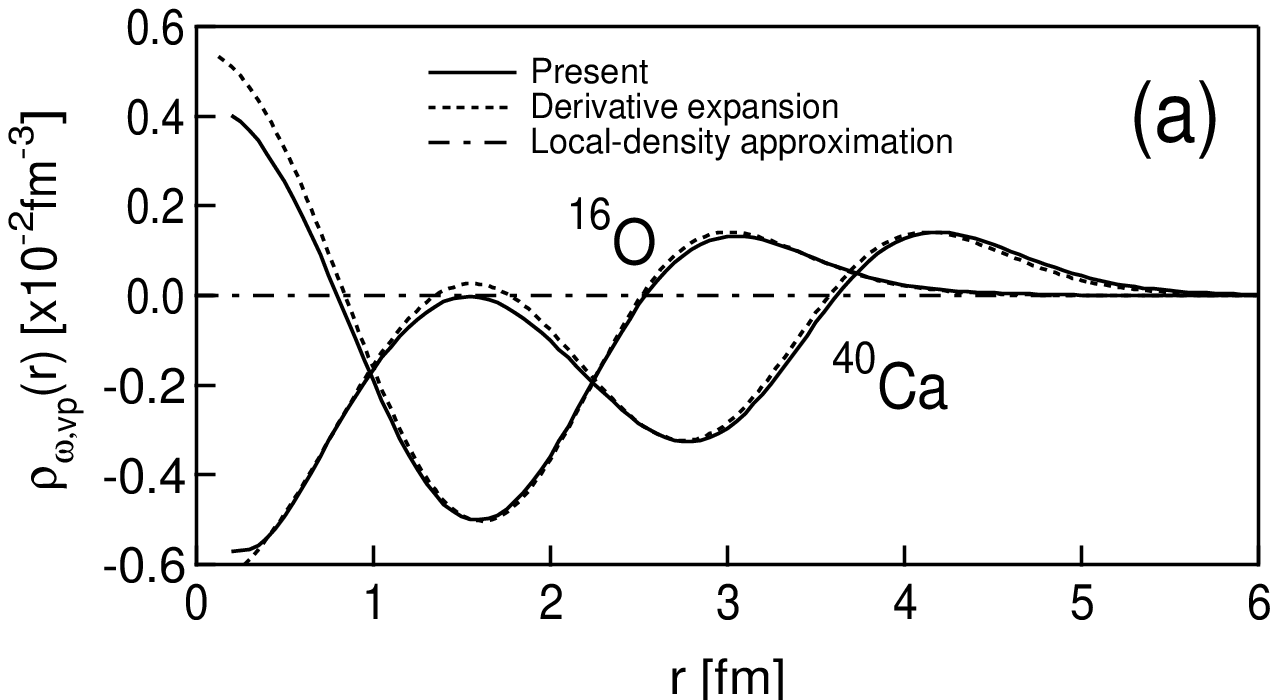}
\includegraphics[width=7.0cm,clip]{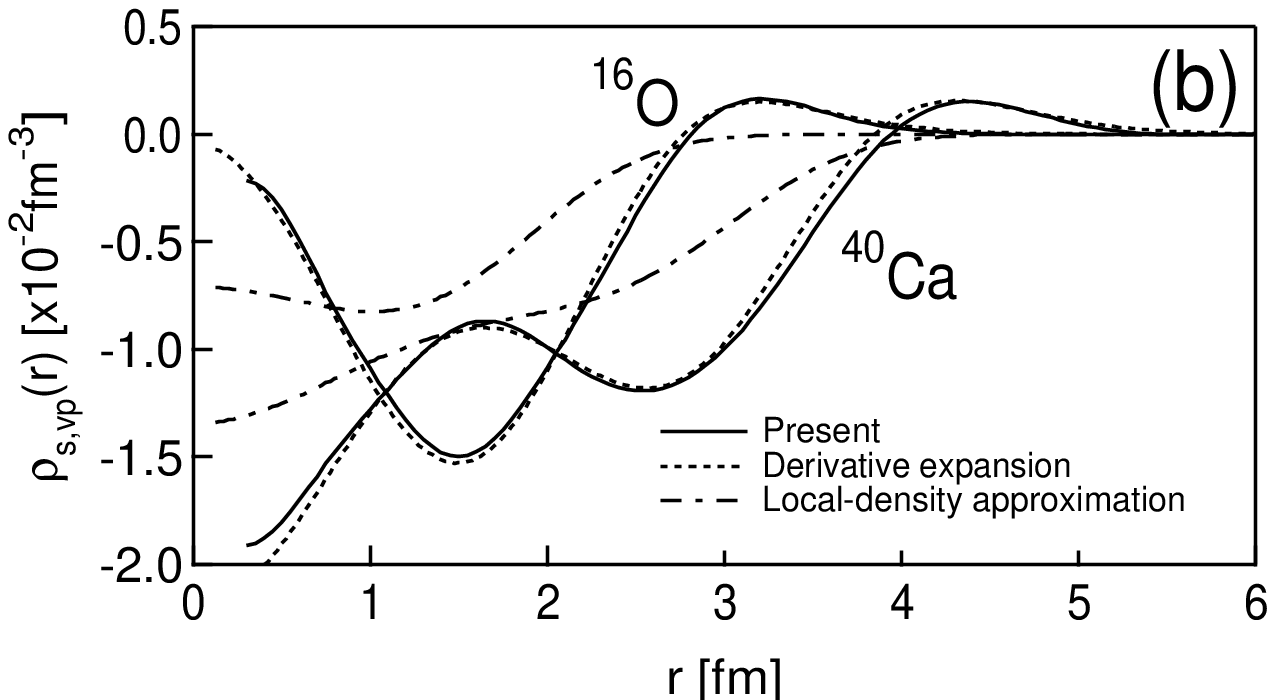}
\caption{\label{fig10}
Vacuum correction for (a) baryon and (b) scalar densities.}
\end{figure}


\begin{thebibliography}{30}

\setlength{\baselineskip}{0.3in}

\bibitem{WA74}
J. D. Walecka, Ann. Phys. (N.Y.) {\bf 83}, 491 (1974);
B. D. Serot and J. D. Walecka,
                Adv. Nucl. Phys. {\bf 16}, 1 (1986);
B. D. Serot and J. D. Walecka,
                J. Mod. Phys. E{\bf 6}, 515 (1997).

\bibitem{KL02}
E. Klempt, F. Bradamante, A. Martin, and J. Richard,
                Phys. Rep. {\bf 368}, 119 (2002).

\bibitem{GR95}
W. Greiner, Heavy Ion Physics {\bf 2}, 23 (1995).

\bibitem{HA04}
A. Haga, Y. Horikawa, Y. Tanaka, and H. Toki,
                Phys. Rev. C{\bf 69}, 044308 (2004).

\bibitem{HO84}
C. J. Horowitz and B. D. Serot,
                Phys. Lett. B {\bf 140}, 181 (1984).

\bibitem{PE86} R. J. Perry,
                Phys. Lett. B {\bf 182}, 269 (1986).

\bibitem{IC88}
S. Ichii, W. Bentz, A. Arima, and T. Suzuki,
                Nucl. Phys. A{\bf 487}, 493 (1988).


\bibitem{WA88} D. A. Wasson,
                Phys. Lett. B {\bf 210}, 41 (1988).

\bibitem{FO89}
W. R. Fox,      Nucl. Phys. A{\bf 495}, 463 (1989).

\bibitem{FU89}
R. J. Furnstahl and C. E. Price,
                Phys. Rev. C{\bf 40}, 1398 (1989);
                Phys. Rev. C{\bf 41}, 1792 (1990).

\bibitem{MA99}
G. Mao, H. St{\"o}cker, and W. Greiner,
                Int. J. Mod. Phys. E{\bf 8}, 389 (1999).

\bibitem{MA03}
G. Mao,         Phys. Rev. C{\bf 67}, 044318 (2003).

\bibitem{GY72}
M. Gyulassy,    Phys. Rev. Lett. {\bf 33}, 921 (1974);
                Phys. Rev. Lett. {\bf 32}, 1393 (1974);
                Nucl. Phys. A{\bf 244}, 497 (1975).

\bibitem{SO88}
G. Soff and P. J. Mohr,
                Phys. Rev. A{\bf 38}, 5066 (1988).

\bibitem{BL90}
P. G. Blunden,
                Phys. Rev. C{\bf 41}, 1851 (1990).

\bibitem{ST97}
I. W. Stewart and P. G. Blunden,
                Phys. Rev. D{\bf 55}, 3742 (1997).

\bibitem{WI56}
E. H. Wichmann and N. M. Kroll,
                Phys. Rev. {\bf 101}, 843 (1956).

\bibitem{UE35}
E. A. Uehling,  Phys. Rev. {\bf 48}, 55 (1935).

\bibitem{RI75}
G. A. Rinker, Jr. and L. Wilets,
                Phys. Rev. A{\bf 12}, 748 (1975).

\bibitem{SU94}
Y. Sugahara and H. Toki,
                Nucl. Phys. A{\bf 579}, 557 (1994).

\bibitem{DATA}
J. H. E. Mattauch, W. Thiele, and A. H. Wapstra,
                Nucl. Phys. {\bf 67}, 1 (1965);
J. W. Negele, Phys. Rev. C{\bf 1}, 1260 (1970);
D. Vautherin and D. M. Brink,
                Phys. Rev. C{\bf 5}, 626 (1972);
H. de Vries, C. W. de Jager, and C. de Vries,
                At. Data Nucl. Data Tables {\bf 36}, 495 (1987).

\bibitem{OG04}
Y. Ogawa, H. Toki, S. Tamenaga, H. Shen, A. Hosaka, S. Sugimoto,
and K. Ikeda,
                Prog. Theore. Phys. {\bf 111}, 75 (2004).

\bibitem{SU942}
Y. Sugahara and H. Toki, Prog. Theor. Phys. {\bf 92}, 803 (1994).


\bibitem{KT91}
K. Tanaka, W. Bentz, A. Arima, and F. Beck,
                Nucl. Phys. A{\bf 528}, 676 (1991).





\end{thebibliography}
\end{document}